# Coupled Ferroelectricity and Phonon Chirality

Xiang-Bin Han,[1†] Cong Yang,[2†] Rui Sun,[3] Xiaotong Zhang,[3] Thuc Mai,[4] Zhengze Xu,[2] Aryan Jouneghaninaseri,[2] Xiaoning Jiang,[2] Rahul Rao,[4] Yi Xia,[5*] Dali Sun,[3*] Jun Liu,[2*] Xiaotong Li[1*]

†The authors contribute equally
1. Department of Chemistry and Organic and Carbon Electronics Laboratories (ORaCEL), North Carolina State University, Raleigh, North Carolina 27695, USA
2. Department of Mechanical and Aerospace Engineering and Organic and Carbon Electronics Laboratories (ORaCEL), North Carolina State University, Raleigh, North Carolina 27695, USA
3. Department of Physics and Astronomy and Organic and Carbon Electronics Laboratories (ORaCEL), North Carolina State University, Raleigh, North Carolina 27695, USA
4. Materials and Manufacturing Directorate, Air Force Research Laboratory, Wright-Patterson AFB, Dayton, Ohio 45433, USA
5. Department of Mechanical and Materials Engineering, Portland State University, Portland, Oregon 97201, USA
Email: yxia@pdx.edu, dsun4@ncsu.edu, jliu38@ncsu.edu, xiaotong_li@ncsu.edu

**Abstract**
The ability to control chirality and chiral phonons offers a route to manipulate the direction of spin and angular-momentum transport. In materials with rigid structural chirality, such as quartz, phonon chirality is fixed by the handedness and cannot be switched. By contrast, ferroelectric materials host a spontaneous polarization that can be reversibly switched by an external electric field. When chirality is coupled to this ferroelectric polarization, it enables electrical switching of crystal chirality and the associated phonon angular momentum, which is compatible with solid-state spintronic architectures, enabling control over chirality-dependent quantum states.[1] Here, we report the experimental demonstration of the coupling between ferroelectricity and phonon chirality in the molecular ferroelectric triglycine sulfate. By electrically switching the crystal chirality, we achieve reversible and device-compatible control of phonon chirality, as revealed by *in situ* time-resolved magneto-optical Kerr effect measurements. The Kerr rotation reverses with electric-field switching, while phonon chirality vanishes in the paraelectric phase and is tunable in the racemic ferroelectric state. Furthermore, density functional theory calculations and circularly polarized Raman spectroscopy further corroborate the opposite circular phonon motions. These results establish an electrically addressable coupling pathway linking ferroelectricity, structural chirality, chiral phonons, and spin, opening a route toward chiral-phonon-enabled spin and phonon control technologies based on ferroelectric materials.

**Introduction**

Chiral phonons are collective crystal lattice vibration modes carrying angular momentum and circular polarization, first proposed in 2015.[2-4] Their existence relies on the non-centrosymmetric structure,[5] and has been theoretically predicted in kagome,[6] moiré superlattices,[7] and materials with broken inversion symmetry.[8-11] Experimentally, chiral phonons have been identified using circularly polarized optical and X-ray spectroscopies,[12-14] time-resolved magneto-optical Kerr effect (TR-MOKE) measurements utilizing the chiral-phonon-activated spin Seebeck effect,[15] circularly polarized Raman[16-17] and terahertz spectroscopy,[18-19] and spin–phonon coupling effects in various chiral or polar materials.[20-21] Chiral phonons have attracted significant attention because heat flow can excite angular-momentum-carrying lattice vibrations, which transfer angular momentum to electronic spins through electron–phonon coupling, thereby enabling spin control in spintronics.[5, 22] Despite this progress, chiral phonons have been observed mainly in materials with rigid structural chirality, where the phonon chirality is fixed by the handedness.[4] The absence of reversible chirality control limits the integration of chiral phonons into functional devices that require deterministic switching of angular momentum, motivating the need for a solid-state platform in which phonon chirality can be reversibly switched. From a device perspective, if an electric field, the natural control parameter in solid-state spintronic devices, can be used to control phonon chirality through electron–phonon coupling, additional control mechanisms that increase device complexity can be avoided.

Ferroelectric materials exhibit a switchable spontaneous polarization under an external electric field and the broken inversion symmetry required by chiral phonons, which provide a perfect platform for exploring symmetry-governed properties, especially relevant to quantum phenomena such as chiral phonons.[23-33] Recent studies have revealed a potential coupling between ferroelectricity and chirality in optically active ferroelectrics,[1] where electric polarization and structural chirality can be coupled for crystals in five out of ten ferroelectricity active polar point groups. This coupling enables the electrical control of chirality, providing a means to modulate a wide range of chirality-related properties, including optical rotation, circular dichroism, circularly polarized luminescence, asymmetric synthesis/catalysis, and chirality-induced spin selectivity.[1, 34-35] In this work, we further demonstrate that electric-field–induced switching of crystal chirality enables direct control of phonon chirality via ferroelectricity, thereby extending the functional scope of ferroelectrics beyond conventional polarization switching toward spin–lattice–chirality interactions.

Several theoretical studies have predicted that ferroelectricity could provide an effective route to manipulate chiral phonons,[36-39] yet experimental realization has remained elusive due to practical challenges. Building on the intrinsic coupling between ferroelectricity and chirality discussed above, here we present the first experimental demonstration of in situ control of phonon chirality in a molecular ferroelectric crystal, triglycine sulfate (TGS).[1] By electrically switching the crystal chirality via an external electric field, we directly observe reversible modulation of phonon chirality using TR-MOKE measurements. Our results establish a sequential coupling pathway in which ferroelectricity controls structural chirality, chirality activates chiral phonons under heat flow, and chiral phonons transfer angular momentum to electronic spins, thereby enabling electrical control of spin (Figure 1a). Such electrical manipulation of spin via chiral phonons opens new opportunities for chiral-phonon-driven spintronic and phononic devices.

**Results and discussion**

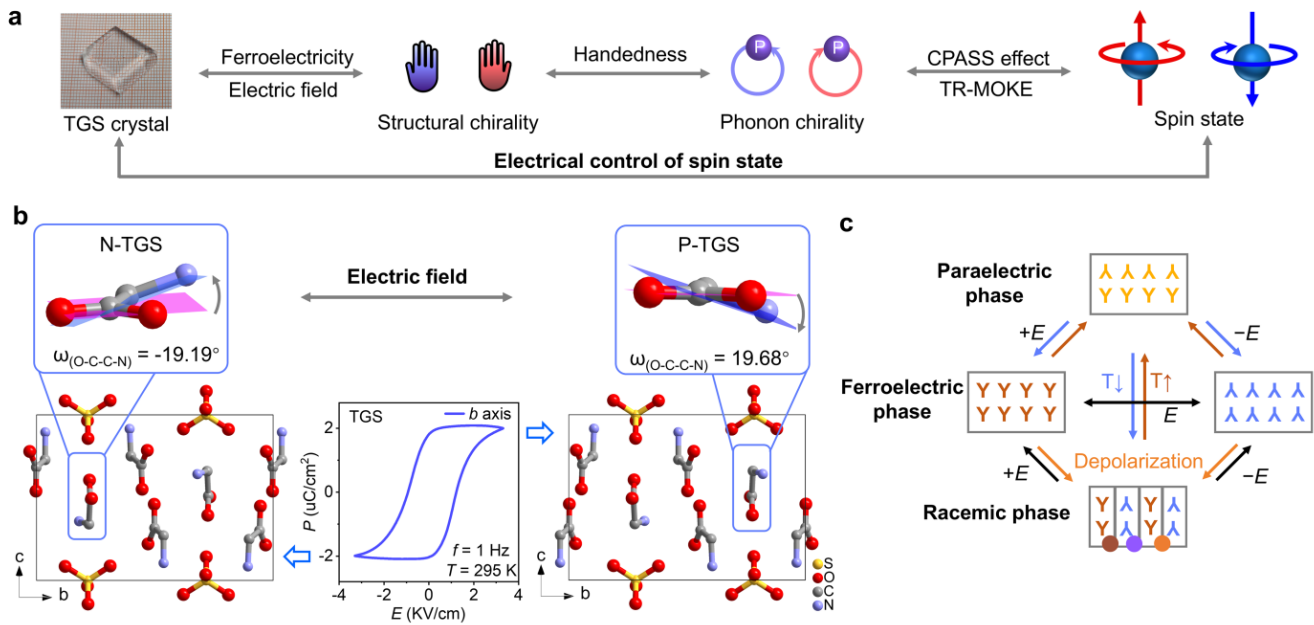

**Figure 1** Schematics for coupled ferroelectricity and chirality in TGS crystal. (a) The coupling chain to achieve the electrical control of spin state between TGS crystal and spin state cont. CPASS: chiral-phonon-activated spin Seebeck effect. (b) Crystal structure of N-TGS exhibits a negative O-C-C-N dihedral angle, which can be switched to P-TGS with a positive dihedral angle under the electric field. Polarization-electric (*P-E*) field hysteresis loop. (c) Phase transitions of TGS illustrating the evolution of polarization and chirality. The orientation of "Y" denotes the absolute molecular configuration within the crystal and indicates the directions of both chirality and ferroelectric polarization. T↑: heating; T↓: cooling, E: electric field.

Ferroelectricity depends on the presence of polarity, since it inherently breaks inversion symmetry. Among the 10 polar point groups that allow ferroelectricity, 5 are chiral that provide the potential to switch chirality and chiral phonons by applying external electric field. To investigate the coupling between ferroelectricity and chiral phonons, we study a classic molecular ferroelectric compound, TGS, because of several practical advantages in device fabrication (see below). TGS is composed of one neutral glycine molecule with a dihedral angle, two protonated glycine cations that are coplanar, and one sulfate anion. It crystallizes in the space-group type, $P2_1$, where structural chirality and ferroelectric polarization are intrinsically coupled, arising from the torsional distortion of the neutral, non-coplanar glycine molecule together with the cooperative polar displacement of the sulfate anions along the *b* axis. The electric field induces a change in the O–C–C–N dihedral angle from positive (+19.68°) to negative (–19.19°) (Figure 1a-b). The switchable polarity is demonstrated by *P–E* loop measurements (Figure 1c), while the chirality flip has been confirmed through single-crystal X-ray diffraction,[1] where the P-TGS shows the opposite dihedral angle compared to the N-TGS. It is also confirmed by optical rotatory dispersion[40] and circular dichroism[41] measurements.

The TGS crystal exhibits four distinct phases with different symmetry and chirality/polarity states during its phase transition process, which can be interconverted under specific external conditions (Figure 1d).[41] These phase transitions provide three practical routes for phonon control: (i) temperature tuning across the ferroelectric–paraelectric transition, (ii) electric-field manipulation within the ferroelectric phase, and (iii) modulation of net chirality in different ferroelectric domains in the racemic phase. At elevated temperatures above 49 °C, TGS adopts a centrosymmetric, non-polar paraelectric phase ($P2_1/m$, Table S1). Upon cooling under an applied electric field, it transforms into one of two ferroelectric phases with opposite polarization, homochirality, and optical activity, which are switchable by reversing the electric field direction. In contrast, if cooled without an external field, a racemic ferroelectric phase forms, consisting of an equal mixture of left- and right-handed chiral domains. These four states—non-polar paraelectric phase, left/right-handed ferroelectric phase, and racemic ferroelectric phase—are interrelated through temperature changes and electric-field-driven switching or depolarization processes (Figure 1d).

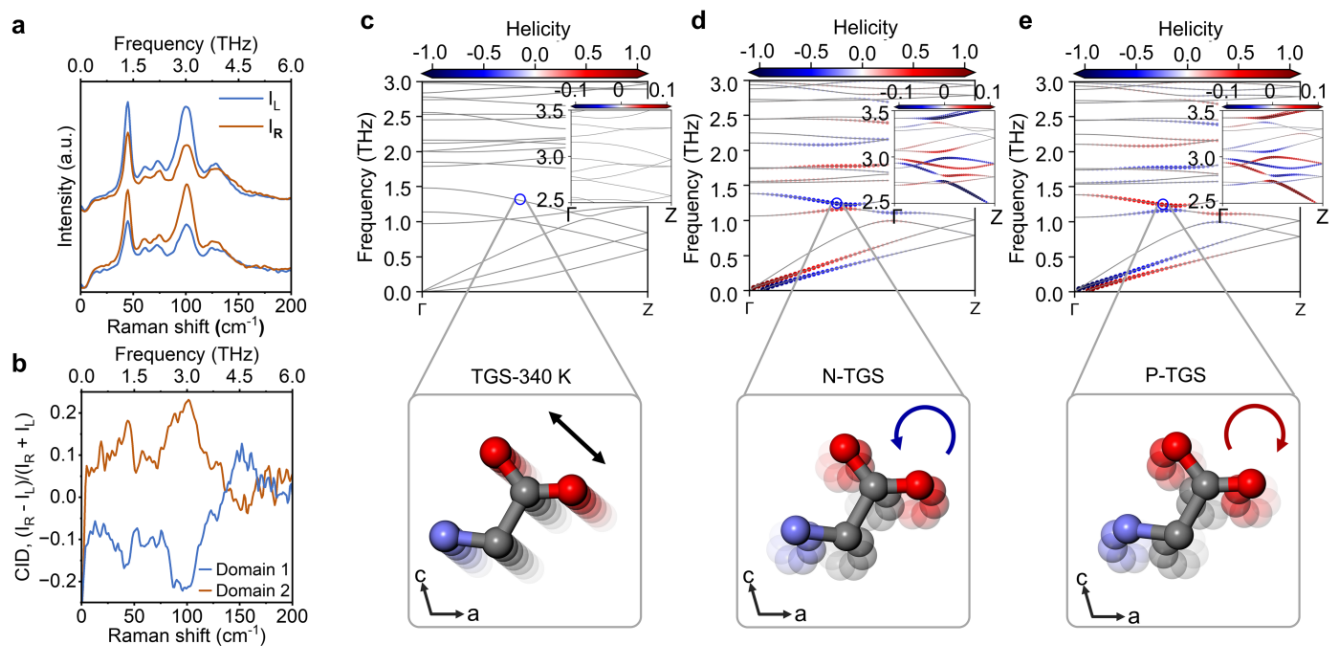

**Figure 2** Chiral phonons in TGS revealed by circularly polarized Raman spectroscopy and first-principles calculations. (a) Circularly polarized Raman spectra of TGS in two different ferroelectric domains. (b) Corresponding circular intensity difference (CID) spectra from the two TGS domains; $I_R$ and $I_L$: denote Raman intensities measured under opposite circular polarization configurations. (c-e) Calculated phonon helicity along the Γ–Z direction and representative phonon modes for (c) paraelectric phase TGS at 340 K, and ferroelectric phase (d) N-TGS, and (e) P-TGS. Red and blue colors indicate phonon modes with opposite helicities, insets show the magnified helicity in 2.5 – 3.5 THz range. The lower panels show displacement trajectories of a representative glycine molecule along the $b$ axis. The color scale of the calculated helicity does not correspond to the magnitude of the experimental CID.

Cross-circularly polarized Raman spectra were collected on racemic TGS crystals in different domain spots. Five prominent peaks are observed at 45, 61, 74, 103 and 128 cm$^{-1}$, originating mainly from glycine vibrational modes in TGS (Figure 2a).[42-44] The differential absorption and scattering is more apparent in the circular intensity differences (Figure 2b, Figure S1), CID=$(I_R-I_L)/(I_R+I_L)$, where $I_R$ and $I_L$ denote Raman intensities measured under opposite circular polarization configurations, $I_R- I_L$ is the Raman optical activity, and the CID is its strength. Pronounced bisignate features are observed for low-frequency optical phonon modes, with a maximum CID of 0.23 at 103 cm$^{-1}$ (~3 THz) (details in Supplementary Discussion 4). These CID values exceed those reported for typical organic chiral materials (~10$^{-3}$)[45] and are comparable to recent reports in two-dimensional materials and hybrid organic–inorganic perovskites.[46-49] The large CIDs indicate a strong coupling between glycine-related optical phonons and structural chirality in TGS. Fits to the Raman peaks yield full widths at half maximum of 7.2 and 17.8 cm$^{-1}$ for the phonon modes near 1 and 3 THz, corresponding to phonon lifetimes of approximately 0.74 and 0.30 ps, respectively, which are consistent with expectations for short-lived optical phonons.[50]

To further verify the phonon vibration modes in the paraelectric and two equivalent ferroelectric phases, we performed first-principles calculations of the phonon dispersion based on DFT[51-52] using the Vienna Ab initio Simulation Package (VASP).[53-56] The simulations were conducted for the centrosymmetric paraelectric phase (TGS–340 K) and for the two ferroelectric phases (N-TGS and P-TGS). For N-TGS and P-TGS, previously reported experimental structures were used as the input atomic coordinates.[1] For the paraelectric phase, the structure of TGS at 340 K was determined experimentally, and the twofold-disordered NH$_3$ groups were averaged into a single position within the glycine molecular plane to generate an input model. This treatment avoided artificial atom-number doubling and occupation splitting, which cannot be properly represented as ordered structure in DFT calculation. After symmetry-constrained structural optimization, the dihedral angle of the glycine molecules increased slightly by about 3° (Table S2).

For structures in point group 2 symmetry, the unique axis, $b$ axis, serves as both the chiral and polar axis, responsible for the generation of chirality and polarity. Accordingly, the phonon dispersion along the Γ–Z direction (parallel to the real-space $b$ axis) was selected for TGS–340 K, N-TGS, and P-TGS (Figures 2c–e), to show the dispersions along with the phonon angular momentum (PAM), where opposite PAM is indicated by red or blue color. In the paraelectric phase (TGS–340 K), the computed phonon helicities are zero, indicating the absence of phonon chirality. In contrast, N-TGS and P-TGS exhibit clear but opposite chiral phonons modes at corresponding points in the Brillouin zone (Figure 2d-e), where blue and red signals denote counterclockwise and clockwise molecular rotations, respectively. For a representative glycine molecule, the vibrational motion is linear in paraelectric phase, whereas in N-TGS it evolves into a two-dimensional counterclockwise rotational motion within the $ac$ plane, and reverses to a clockwise rotation in P-TGS. At frequency around ~1.3 and 3 THz, where the Raman CID value peaks, opposite PAM is also predicted by the calculations (insets in Figures 2c–e). A representative phonon mode in this frequency range was visualized using the TSS Physics Visualization of Phonons program (Video S1). In this mode, molecular units in N-TGS and P-TGS undergo circular motions with opposite handedness, while only linear oscillations persist in the paraelectric phase. These results establish chiral optical phonons as the microscopic origin of the Raman optical activity observed in TGS.

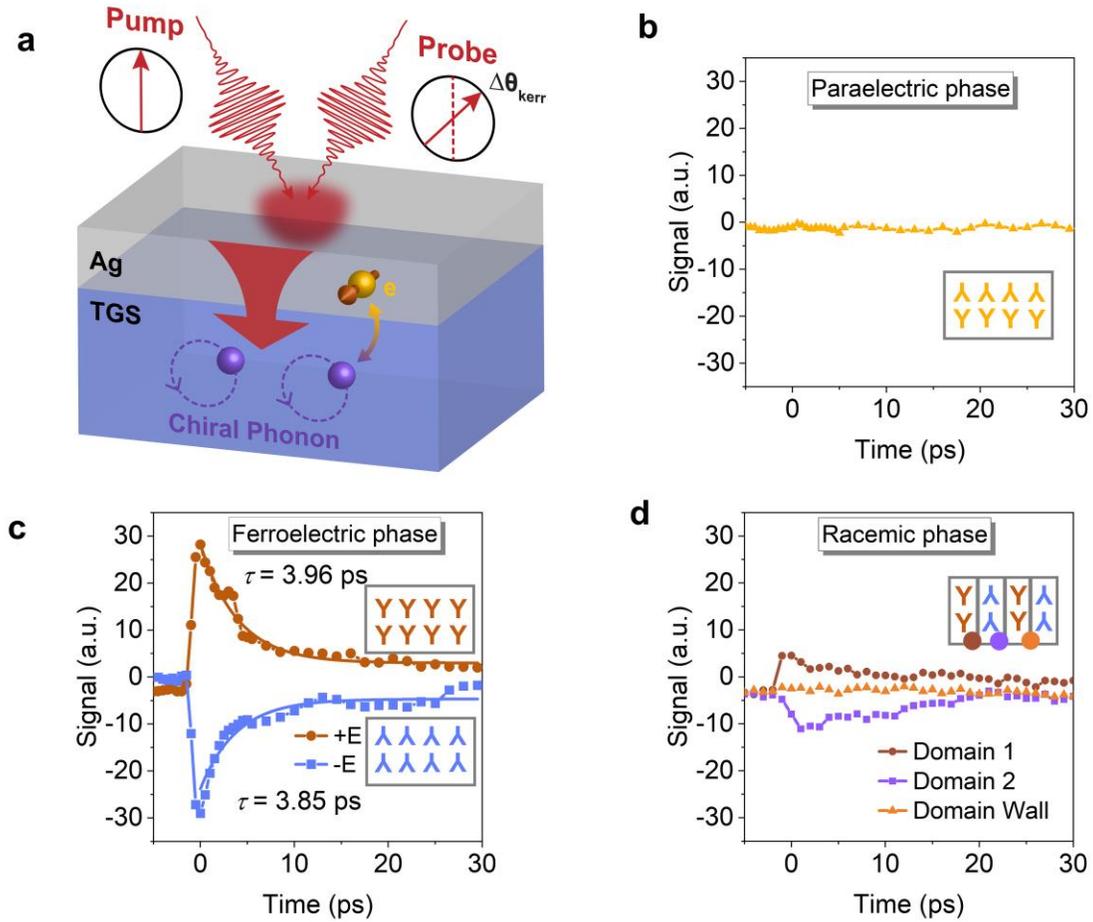

**Figure 3** Chiral phonon signals in TGS. (a) Schematic illustration of the TR-MOKE mechanism. (b) No chiral phonon signal is observed in paraelectric phase. (c) Switchable chiral phonon signals in TGS under positive/negative poling, the solid lines are the nonlinear fitting curves. (d) Domain-dependent chiral phonon signals in racemic TGS.

Having established the presence and microscopic origin of chiral phonons in TGS by circularly polarized Raman spectroscopy and first-principles calculations, we next examine their electric-field-induced switching. Since electrode coverage precludes Raman measurements with the electric field, we chose to use TR-MOKE measurements, which allow the use of metal electrodes that are compatible with electric-field application. TR-MOKE probes transient spin accumulation through ultrafast polarization changes in reflected light, providing a sensitive readout of phonon angular momentum. In our experiment, chiral phonons were probed indirectly via Ag-surface Kerr rotation (Figure 3a), using a 783 nm pump–probe configuration (Figure S2), enabled by the chiral-phonon-activated spin Seebeck effect.[15, 57] Unlike the conventional spin Seebeck effect that relies on a ferromagnetic insulator, chiral phonons in a non-magnetic chiral crystal generate spin-polarized currents under a thermal gradient. This mechanism, demonstrated experimentally[15] and supported by theoretical analyses,[22, 58] allows angular momentum carried by chiral phonons to be transferred to conduction electrons in the adjacent metal layer, producing a measurable Kerr rotation.

TGS is an ideal molecular crystal system for investigating the coupling between ferroelectricity and chiral phonons using the TR-MOKE setup, because of several practical advantages. These include the ease of growing large single crystals via solution processing, and the presence of a natural cleavage plane that exposes the (020) facet, which perpendicular to both the polar and chiral axes (Figure S3). Furthermore, TGS allows for the preparation of surfaces with sub-nanometer roughness, both before and after metal deposition, resulting in mirror-like surfaces suitable for optical measurements, which require mirror-reflection conditions. Atomic force microscopy reveals a root-mean-square roughness of 138.8 pm on the as-cleaved surface, which slightly increases to 331.9 pm after depositing a 50 nm Ag layer (Figure S4). Additionally, the domain size of TGS reaches the sub-millimeter scale,[59-60] significantly larger than the 6 μm diameter of the probe laser beam.

Upon pump excitation of the device (Figure S5, details in Supplementary Discussion 5) on 50 nm Ag layer, heat is rapidly injected across the Ag/TGS interface into the underlying 1 mm-thick TGS crystal, establishing a transient thermal gradient. This thermal gradient drives the generation and propagation of chiral phonons carrying net angular momentum toward the Ag/TGS interface, where their angular momentum is transferred to conduction electrons, leading to spin accumulation in the Ag layer. Owing to the long spin diffusion length in Ag (357 nm),[61] the spin-polarized electrons diffuse to the Ag surface, resulting in a Kerr rotation signal detected by the time-delayed probe beam. Prior to the measurement, the thermal conductivity of TGS was evaluated to determine an appropriate incident light power (5 mW), ensuring that surface damage and temperature-induced phase transition to the paraelectric phase ($T_C$ ~ 49 °C) were avoided. The measured thermal conductivity was 0.54 ± 0.07 W/(m·K), consistent with previously reported values (Figure S6).[62]

In the paraelectric phase, no Kerr signal is observed (Figure 3b), consistent with the centrosymmetric structure and the absence of chiral phonons. In contrast, clear Kerr rotation signals are detected in both ferroelectric phases of TGS with opposite chirality and polarity during TR-MOKE measurements (Figure 3c). These signals arise from spin accumulation induced by the angular momentum carried by chiral phonons. The Kerr rotation exhibits opposite signs, denoted as $+\Delta\theta_{Kerr}$ and $-\Delta\theta_{Kerr}$, corresponding to opposite chirality states, while the peak magnitude reflects the absolute Kerr rotation angle. The signals decay rapidly to the baseline within ~10 ps, attributable to ultrafast spin relaxation. Nonlinear fitting yields comparable spin decay times of ~4 ps for positively and negatively poled TGS (3.96 ps and 3.85 ps, respectively). The measurements are highly reproducible, with three cycles yielding repeatable TR-MOKE responses (Figure S7).

When cooled without an external electric field, a racemic phase forms, consisting of regions with equal and opposite optical activity that cancel each other out, as confirmed by circular dichroism measurements.[41] Regions of the same chirality/polarity correspond to ferroelectric domains, while the domain walls separating them are dynamic, gradually expanding over time from an initial micrometer-scale width.[59] In this state, three types of Kerr rotation signals can be detected (Figure 3d): positive and negative signals arise from domains with net chirality, while a flat curve appears when the laser spot is located at the center of a domain wall, where the net chirality vanishes. At such boundaries, left- and right-handed chirality coexist, forming a locally racemic, pseudo-centrosymmetric state without chiral phonon signal. The Kerr rotation amplitude in the racemic crystal is approximately one-third of that observed under fully poled conditions. This reduction reflects the partial cancellation of chirality between adjacent domains ([Figure 3d](#)), indicating cancellation of opposite chiral contributions and a residual net chirality weaker than that in the homochiral state.

To exclude the contribution of thermally induced effects in the TR-MOKE signal, control measurements were conducted using time-domain thermoreflectance (TDTR), a technique widely used to probe thermal transport and transient heating at the nanoscale.[63] In TDTR, the pump beam induces a local temperature rise, and the corresponding change in reflectance is monitored by a delayed probe beam, allowing thermal diffusion and conductivity to be quantitatively evaluated. In the TDTR control measurements, the thermal signals, with applied positive or negative electric field, exhibited the same temporal behavior and direction ([Figure S8](#)). The similar decay rates indicate consistent thermal diffusivity, while the invariant peak amplitude rules out changes in energy absorption. This result confirms that the Kerr signal observed in the TR-MOKE experiment does not originate from purely thermal effects. Instead, it arises from spin dynamics induced by optical chiral phonons carrying angular momentum, further excluding the influence of thermally driven processes.

**Conclusion**

We demonstrate that chiral phonons in the ferroelectric crystal triglycine sulfate (TGS) can be reversibly switched by an external electric field, as evidenced by Kerr rotation signals detected at the Ag surface. By electrically switching the crystal chirality, the phonon chirality is correspondingly reversed, while the paraelectric phase completely suppresses chiral phonons. In the racemic ferroelectric state, the phonon chirality can be continuously tuned by controlling the net crystal chirality. The reversal of phonon chirality is further corroborated by first-principles calculations and circularly polarized Raman spectroscopy, which reveal opposite collective molecular rotations in N- and P-TGS. Overall, this work establishes ferroelectricity as an effective control knob for activating and manipulating chiral phonons and their associated circular polarization, enabling electrical control of spin. By advancing chiral phonons from static observation to active electrical manipulation, our results open new opportunities for chiral-phonon-based phononic and spintronic devices.

**Acknowledgements**

X. Li acknowledges the support by the UNC Research Opportunity Initiative and startup funds from NC State University. J. Liu and D.S. acknowledge the support by the Air Force Office of Scientific Research (ADOSR), Multidisciplinary University Research Initiatives (MURI) Program under award number FA9550-23-1-0311. J. Liu also acknowledge the financial support from the National Science Foundation with the award number DMR-2521953. Z.X. and X.J. acknowledge the support by ONR under Grant # N00014-21-1-2058. T.M. and R.R. acknowledge the support of AFOSR grant number RX23COR003. Y.X. acknowledges the support from the U.S. National Science Foundation through Awards No. DMR-2317008 and No. CBET-2445361 and the computing resources provided by Bridges2 at Pittsburgh Supercomputing Center (PSC) through allocation mat220008p from the Advanced Cyber-infrastructure Coordination Ecosystem: Services and Support (ACCESS) program, which is supported by National Science Foundation Grants No. 2138259, No. 2138286, No. 2138307, No. 2137603, and No. 2138296.

The authors thank Ben Hines, Subhrangsu Mukherjee, and Harald Ade for their assistance in depositing the Ag layer.

**Author contributions**
X.-B. H. conceived the research concept together with D. S., J. L., and X. L., and prepared the devices as well as wrote the manuscript. C. Y. and A. J. performed the TR-MOKE experiments, and J. L. provided the experimental instrumentation. R. S. assisted with metal deposition, and D. S. provided the corresponding equipment. X. Z. characterized the crystal and device surface roughness using AFM, supported by D. S. who provided the instrument. Z. X. assisted in measuring the $P$–$E$ loop, and X. J. provided the measurement setup. Y. X. carried out the DFT calculations. T. M. and R. R. conducted the circularly polarized Raman measurements. D. S., J. L., and X. Li. supervised the overall project. All authors discussed the results and contributed to revising the manuscript.

**Competing financial interests**
The authors declare no competing financial interests.

**Methods**
**Materials.** Sulfuric acid (98%) and glycine (99%) were obtained from Sigma-Aldrich and used without additional purification. All reagents were of analytical grade and employed directly in the experiments under ambient laboratory conditions unless otherwise specified.

**Device fabrication.** Bulk TGS crystals were cut along the natural cleavage plane perpendicular to the b axis to obtain plate-shaped samples. Powder X-ray diffraction (PXRD) confirmed that the exposed crystal facet was oriented along the b axis, as evidenced by the (020), (040), and (060) diffraction peaks (Figure S3). For device fabrication, a 50 nm Ag layer was deposited on one surface of the TGS crystal after fixing the crystal plate onto a glass substrate. The surface roughness of the Ag-coated side was characterized by AFM (Figure S4). Conductive silver paste was then applied to the opposite surface of the crystal to serve as the counter electrode, and the crystal was attached to a second glass plate using silver paste. Copper wires were connected to both electrodes using silver paste to form an electrical circuit for applying an external electric field. A small amount of silver paste was also applied to partially cover the Ag layer to ensure reliable electrical contact. Kapton tape was used to mechanically secure the copper wires and prevent electrode detachment during measurements (Figure S5a).

The final device for TR-MOKE measurements consists of a TGS crystal sandwiched between an Ag metal layer and a silver paste electrode, mounted on a glass substrate. For thermal integration, the device was sequentially assembled with thermally conductive tape, a resistive heater, and Kapton tape to ensure stable mechanical and thermal contact with the sample holder. Both pump and probe laser beams were focused onto the Ag-coated surface during optical measurements.

**Thermal control.** For temperature control, a resistive heater was mounted onto the optical holder using double-sided Kapton tape. Thermally conductive double-sided tape was used to ensure efficient thermal contact between the heater, the glass substrate, and the TGS crystal. A resistance temperature detector was attached near the sample using the same thermally conductive tape to monitor the sample temperature during measurements (Figure S5b).

**Chirality switching by paraelectric–ferroelectric poling.** In principle, the Kerr signal polarity can be switched by reversing the chirality of TGS under an applied electric field. However, complete chirality reversal is difficult to achieve within a short timescale in the ferroelectric phase, as observed in *P–E* loop measurements. This is due to the thermally driven and inherently slow nature of molecular reorientation or angular adjustments. To overcome this limitation, we employed a paraelectric-to-ferroelectric poling method, in which the crystal is heated into the paraelectric phase and subsequently cooled under a positive or negative electric field. This procedure enables full chirality switching and promotes the growth of large ferroelectric domains.

# Coupled Ferroelectricity and Phonon Chirality

Xiang-Bin Han,[1†] Cong Yang,[2†] Rui Sun,[3] Xiaotong Zhang,[3] Thuc Mai,[4] Zhengze Xu,[2] Aryan Jouneghaninaseri,[2] Xiaoning Jiang,[2] Rahul Rao,[4] Yi Xia,[5*] Dali Sun,[3*] Jun Liu,[2*] Xiaotong Li[1*]

†The authors contribute equally
1. Department of Chemistry and Organic and Carbon Electronics Laboratories (ORaCEL), North Carolina State University, Raleigh, North Carolina 27695, USA
2. Department of Mechanical and Aerospace Engineering and Organic and Carbon Electronics Laboratories (ORaCEL), North Carolina State University, Raleigh, North Carolina 27695, USA
3. Department of Physics and Astronomy and Organic and Carbon Electronics Laboratories (ORaCEL), North Carolina State University, Raleigh, North Carolina 27695, USA
4. Materials and Manufacturing Directorate, Air Force Research Laboratory, Wright-Patterson AFB, Dayton, Ohio 45433, USA
5. Department of Mechanical and Materials Engineering, Portland State University, Portland, Oregon 97201, USA
Email: yxia@pdx.edu, dsun4@ncsu.edu, jliu38@ncsu.edu, xiaotong_li@ncsu.edu

1. Instruments

**Single Crystal X-Ray Diffraction.** Crystallographic data were collected on XtaLAB Synergy R, DW system, and HyPix, with a Hybrid Pixel Array Detector. The X-ray source is Rigaku (Mo) X-ray Source. Data was collected, refined, and reduced with CrysAlisPro 1.171.40.84a (Rigaku OD, 2020). The structures were solved directly using the SHELXL-2019/3 software package. All the non-hydrogen atoms were refined anisotropically, and the positions of hydrogen atoms were generated geometrically. CCDC: 2493177.

**PXRD.** The crystal facet indices were measured using a MiniFlex600-C (Rigaku) diffractometer equipped with a Cu Kα X-ray source, operated at 40 kV and 15 mA.

**Polarization–electric field loop measurement.** The polarization–electric field (P–E) hysteresis loops were recorded using an aixACCT TF Analyzer 2000 (aixACCT Systems GmbH, Aachen, Germany). A TGS crystal coated with silver paste was mounted in a BULK/CMA sample holder to apply the electric field. The P-E is measured with E//[020] direction at 1 Hz.

**Atomic force microscopy (AFM).** Crystal surface roughness was measured using an Asylum MFP-3D atomic force microscope (Asylum Research, Oxford Instruments).

***In-situ* temperature control setup.** The metal-ceramic heater (HT24S, square shape, 24 W, 20 mm×20 mm) and the 100 Ω resistance temperature detector (TH100PT) were controlled by a dual-channel temperature controller (TC300B) using a 6-pin, 3.0 m male-to-male Hirose connector cable (HR10CAB1).

The heater and temperature sensor were attached to a glass substrate using thermally conductive double-sided tape (TCDT1, 1" × 48"). All components were purchased from Thorlabs.

**Voltage source.** A Keithley 236 source measure unit was used as the DC power supply to apply an electric field to the TGS crystal for switching its polarization state. The applied voltage ranged from 400 to 800 V, depending on the crystal thickness, which typically varied between 0.7 and 1.2 mm.

**Metal deposition.** The Ag layer was deposited by thermal evaporation using a thermal evaporation chamber located inside a nitrogen glove box, equipped with an SQC-310C deposition controller. Other metals (Cu, Al) used in this work were deposited by electron-beam evaporation using a deposition system from Angstrom Engineering.

**Time-resolved magneto-optical Kerr effect (TR-MOKE)/ time-domain thermoreflectance (TDTR) setup.** The time-resolved magneto-optical Kerr effect (TR-MOKE)/TDTR measurements were performed using a home-built femtosecond pump–probe system based on a Nd:YVO-pumped Ti:Sapphire laser. The laser output was split into pump and probe beams. The pump beam was focused onto the sample to excite the system, while the probe beam passed through an electro-optic modulator (EOM, 1–10 MHz) and a mechanical delay stage to control the temporal delay between pump and probe pulses. After reflection from the sample, the probe beam carried the Kerr rotation signal associated with the magneto-optical response. The reflected probe passed through polarization optics consisting of a half-wave plate and a Wollaston prism, and the differential signal was detected using a balanced photodetector. Short-pass and long-pass optical filters were employed to separate the pump and probe beams and suppress background signals.

## 2. Bulk crystal growth

**TGS**: Glycine (0.1 M, 7.507 g) and $H_2SO_4$ (98 wt%, 0.033 M, 3.3356 g) were dissolved in 30 mL of water. The resulting solution was then evaporated at room temperature to obtain seed crystals. Selected high-quality seed crystals were subsequently transferred into 200 mL of a glycine–$H_2SO_4$ solution with same concentration to grow large-sized crystals.[1-2]

## 3. A. First-principles calculation of phonons in N-TGS, P-TGS, and TGS-340 K

The atomic coordinates of CIF file used for N-TGS, P-TGS is previous reported with CCDC number of 2343706 and 2343707. The TGS-340 K is used the new reported structure in this work with CCDC number of 2493177. We conducted first-principles simulations based on density functional (DFT) theory[3] to calculate the harmonic phonon dispersions for NTGS, P-TGS, and TGS-340K using the Vienna Ab initio Simulation Package (VASP).[4–7] Our computational approach encompassed structure relaxation and self-consistent DFT calculations. We employed the projector-augmented wave (PAW) method[8] in combination with the vdW-DF2 exchange-correlation functional[9–12] to account for van der Waals interactions. For structure relaxations of the primitive cells of N-TGS. P-TGS, and TGS- 340 K, we utilized Gamma-centered k-point meshes with a minimum spacing of 0.3° $A^{-1}$ between k points. Throughout the study, we maintained a kinetic energy cutoff of 520 eV. To ensure computational accuracy, we set stringent convergence criteria: $10^{-3}$ eV/° A and $10^{-8}$ eV for force and energy, respectively. To compute harmonic phonon dispersions, we employed the Phonopy software package[11]. Our calculations used a 2×1×1 supercell containing 148 atoms in total with the same stringent convergence criteria for energy as the structural relaxation.

We observe that the computed phonon dispersion curves exhibit negative (imaginary) frequencies, indicating dynamical instabilities in the structure. For N/P-TGS, the instabilities near the zone center along the G–Z direction may be attributed to the relatively small real space supercell used for phonon calculation and could potentially be resolved by increasing the supercell size. In contrast, the broader spectrum of negative modes observed along the Z–D–B–Γ-A–E–Z directions may be intrinsic and could be mitigated at finite temperatures due to anharmonic effects. However, we note that the presence of negative phonon frequencies does not affect the calculated phonon helicity, which is fundamentally determined by the crystal symmetry.

**B. Calculation of phonon angular momentum and helicity**

We calculated the mode-decomposed (wave vector q and phonon branch ν) phonon angular momentum $l_{q,\nu}$ using the equation derived by Zhang et al.[14]

$$l_{q,\nu}^\alpha = \hbar \epsilon_{q,\nu}^\dagger M_\alpha \epsilon_{q,\nu} \qquad (1)$$

where the matrix Mα is the tensor product of the unit matrix and the generator of SO(3) rotation for a unit cell with N atoms

$$M_\alpha = I_{N \times N} \otimes \begin{pmatrix} 0 & -i\varepsilon_{\alpha\beta\gamma} \\ -i\varepsilon_{\alpha\gamma\beta} & 0 \end{pmatrix}, \alpha, \beta, \gamma \in \{x, y, z\} \qquad (2)$$

where $\varepsilon_{\alpha\gamma\beta}$ is the Levi-Civita epsilon tensor. Phonon helicity describes the chirality of the phonon mode in terms of phonon angular momentum $l_{q,\nu}$ and the propagation direction q and is defined as[15]

$$h_{q,\nu} = q \times l_{q,\nu} \qquad (3)$$

**4. Circularly polarized Raman**

Room temperature circularly polarized Raman spectra were collected in a Renishaw inVia Raman microscope. Our instrument is outfitted with a low-frequency module (Coherent/Ondax THz Raman probe) that uses fiber optics to couple a 785 nm laser into the objective lens for excitation and to direct the scattered light into the inVia spectrometer. The linearly polarized excitation laser passes through a half and quarter waveplate that changes its polarization to either right- or left-circularly polarized (RCP or LCP, respectively) light. The excitation lase is focused on to the sample through a 50x objective lens. The laser power used for the spectral collection was 1.8 mW, and spectra were collected with a 10 s exposure time and 12 accumulations. The backscattered light from the sample passes through the same waveplates, reversing the polarization state of the scattered light. This produces a cross-circularly polarized configuration (RL or LR). Additionally, the scattered light can also be diverted through a polarizing beamsplitter to obtain co-circularly polarized configuration (RR or LL).

In particular, the modes at 74 and 103 cm$^{-1}$ (2.21 and 3.1 THz, respectively) correspond to librations of non-coplanar glycine about the *a* and *b* axes, respectively, while the lowest frequency peak at 45 cm$^{-1}$ corresponds to in-phase librations of coplanar glycine molecules about the *b* axis.[16] The conformation of non-coplanar glycine is directly responsible for both chirality and ferroelectric polarization in TGS, and has been shown to influence optical activity during ferroelectric switching.[17] While the differences in mode frequencies away from the Brillouin zone center are small enough to preclude observations of Raman peak splitting within our measurement resolution, the low-frequency chiral phonons involving the glycine molecules are strongly affected by circularly polarized light absorption, and thus exhibit high optical activities. Finally, we note that the differences in circularly polarized Raman intensities are not only limited to the low frequencies. As expected, we also observe bisignate peaks in the high frequency region, where the Raman peaks match well with previously reported spectra from TGS.

5. **Choice of electrode metal**

(a) Reactivity series of elements.
The chemical reactivity of common metals follows the order
K > Ca > Na > Mg > Al > Zn > Fe > Ni > Sn > Pb > H > Cu > Bi > Hg > Ag > Pt > Au,
where metals positioned to the left are more chemically active and prone to react with organic functional groups, while those to the right are increasingly inert.

(b) Chemical reactivity between TGS crystals and metals.
Triglycine sulfate (TGS), containing carboxylic acid (–COOH) groups, reacts readily with chemically active metals such as Cu and Al, forming metal carboxylates and releasing hydrogen gas:
$2NH_2CH_2COOH + Cu = Cu^{II}(NH_2CH_2COO)_2 + H_2 (\uparrow)$
$6NH_2CH_2COOH + 2Al = Al^{III}{}_2(NH_2CH_2COO)_6 + 3H_2 (\uparrow)$
In contrast, Ag shows no observable reaction with glycine molecules, highlighting its chemical stability when interfaced with TGS.

(c) Spin diffusion length of typical metals.
Representative spin diffusion lengths are listed for commonly used metals:
Cu: ~500 nm, Al: ~350 nm - 500 nm, and Ag: ~357 nm, indicating that Ag supports the longest spin transport among these candidates.

In the TR-MOKE measurements, silver (Ag) was selected as the top metal on the TGS crystal by considering the metal chemical reactivity and physical spin diffusion length. TGS crystals contain glycine molecules with carboxylic functional groups that can chemically react with various metals. Initially, when copper (Cu) was used, the carboxylic groups readily reacted with Cu atoms— forming copper–carboxylate salts and releasing hydrogen gas. This reaction generated bubbles at the Cu/TGS interface, leading to the chemical exfoliation of the Cu layer from the TGS surface and prohibiting the TR-MOKE measurements. Thus, Ag was selected as the top electrode material through a trial-and-error process for fabricating the single-crystal-based device.

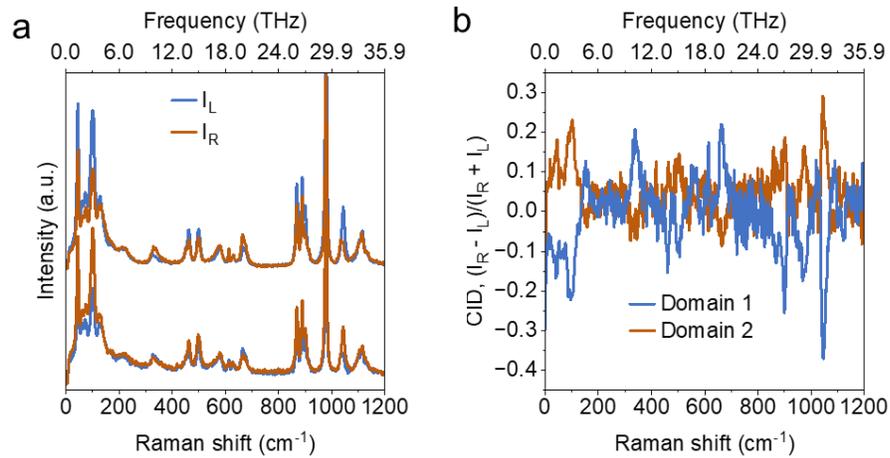

**Figure S1** (a) Circularly polarized Raman spectra from two domains in a racemic TGS crystal. (b) Corresponding circular intensity difference spectra from the two TGS domains.

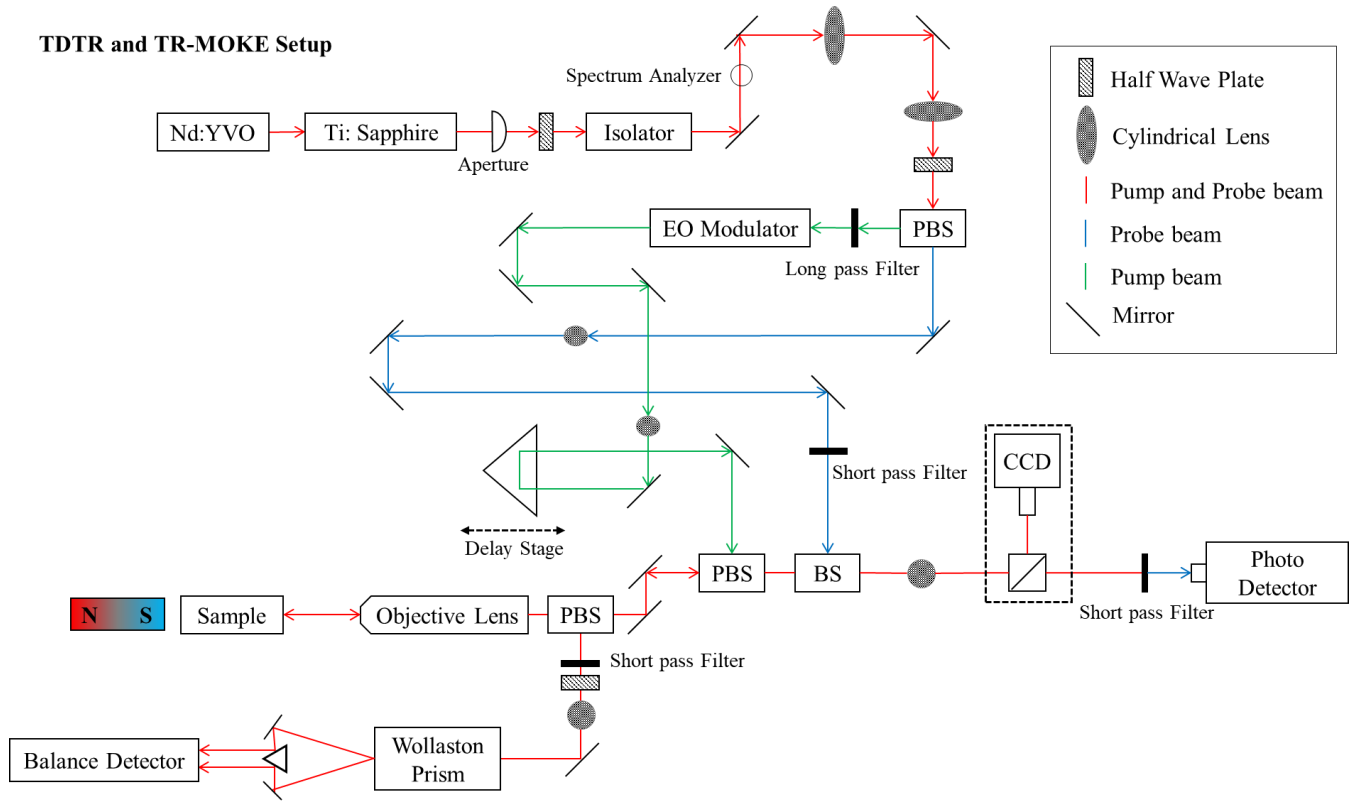

**Figure S2** TR-MOKE and TDTR experiment set up.

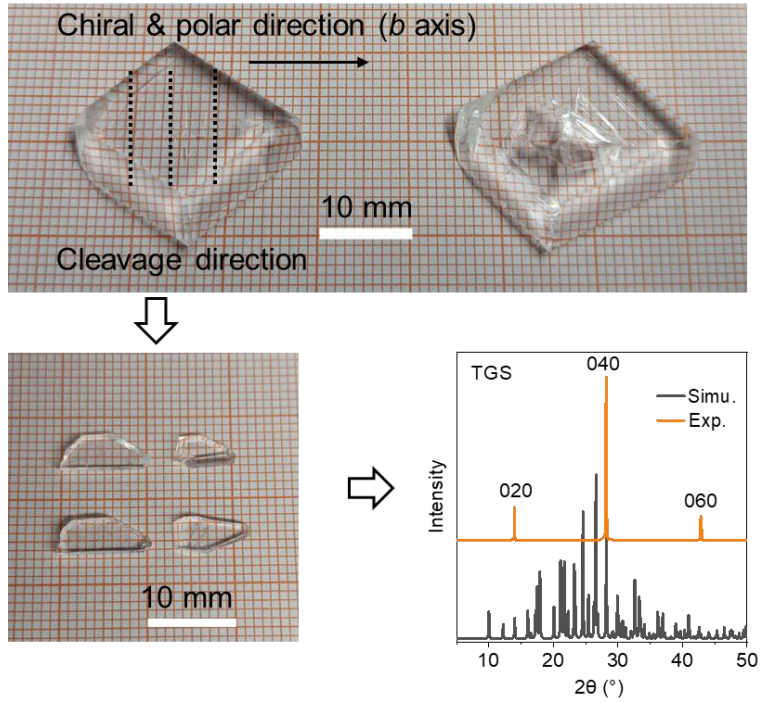

**Figure S3** The as-grown bulk TGS crystal and cleaved crystal plates, with the cleavage direction perpendicular to the *b* axis. The dashed line is the cleavage direction. The crystal facet of cleaved TGS crystal plate is (0k0) plane measured through PXRD. Note: To cleave the crystal, first align the blade parallel to the cleavage plane, then gently tap the back of the blade using the tail of a pair of tweezers to obtain a flake-like crystal.

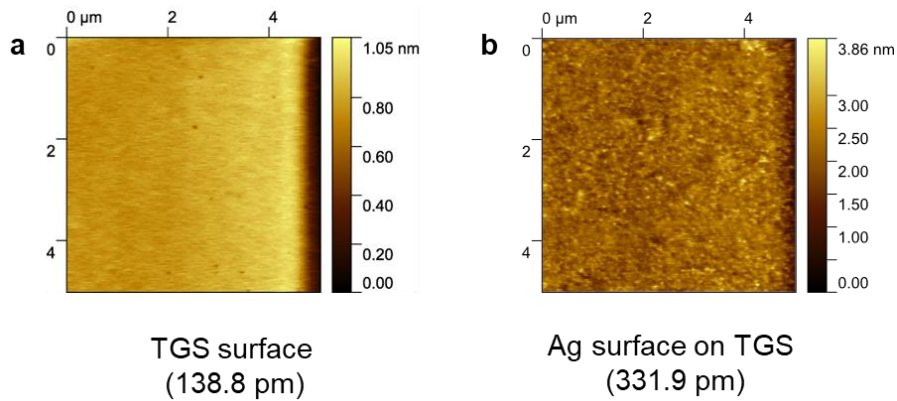

TGS surface
(138.8 pm)

Ag surface on TGS
(331.9 pm)

**Figure S3** (a) AFM image of the (020) crystal facet of TGS. The root mean square (RMS) roughness is 138.8 pm. (b) AFM image showing the morphology of a 50 nm Ag layer deposited on the (020) facet of the TGS crystal. The RMS roughness is 331.9 pm.

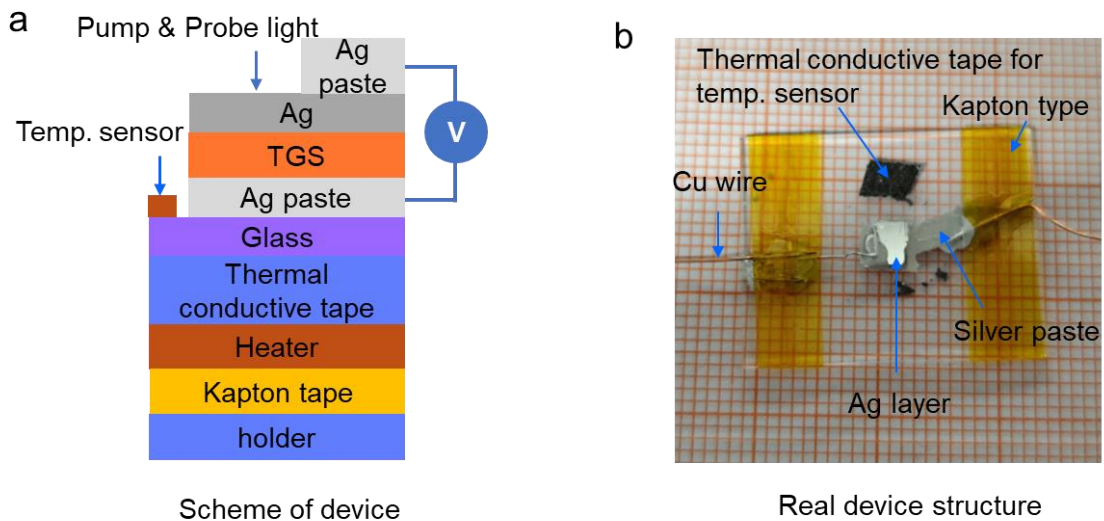

**Figure S5** (a) Schematic illustration of the device structure, and (b) photograph of the actual device.

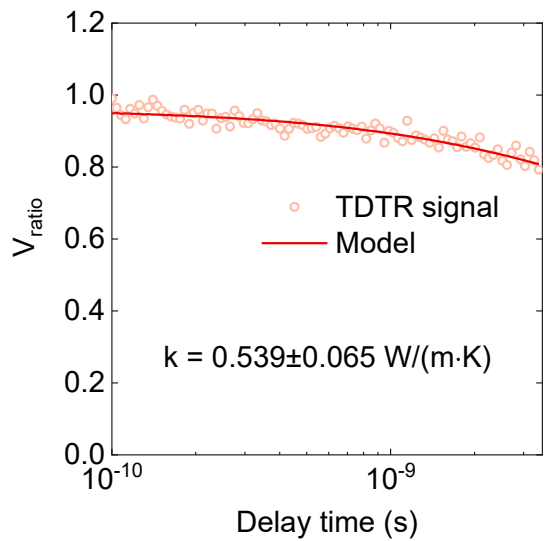

**Figure S6** The thermal conductivity of TGS crystal. Note: In this thermal conductivity measurement, a 50 nm layer of Al was first deposited, and the sample was then placed in a glove box for two months to allow the Al to react with glycine molecules, ultimately forming a stable Al layer suitable for thermal conductivity measurements. Replacing Al with Ag should work in a similar way without aging.

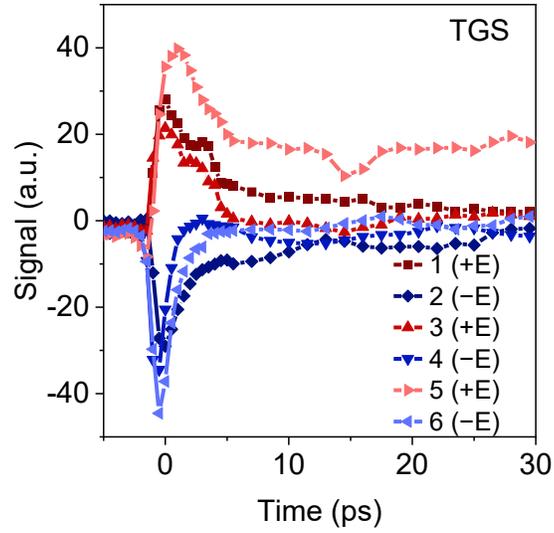

**Figure S7** The chiral phonon signal of TGS under positive and negative electric field for three measurement cycles.

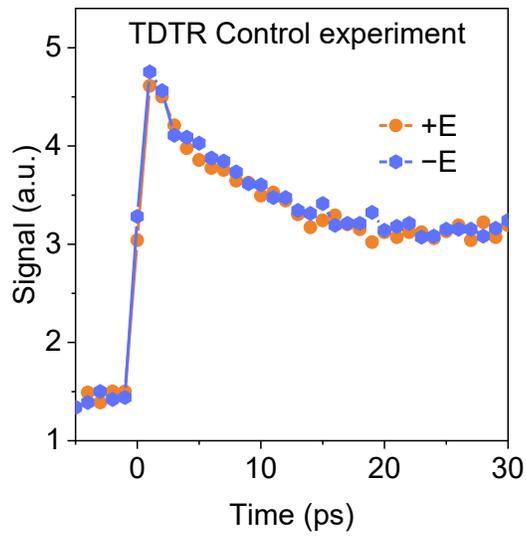

**Figure S8** Time-domain thermoreflectance (TDTR) control experiment of TGS under positive and negative poling electric field.

**Table S1** Crystallographic data and structural refinement details for TGS-340 K.

| Name | TGS |
|---|---|
| CCDC number | 2493177 |
| Formula | $C_6H_{17}N_3O_{10}S$ |
| Formula weight | 323.28 |
| Temperature/K | 340 |
| Crystal system | monoclinic |
| Space group | $P2_1/m$ |
| $a$ [Å] | 5.7518(3) |
| $b$ [Å] | 12.6524(6) |
| $c$ [Å] | 9.1661(4) |
| $\alpha$ [°] | 90 |
| $\beta$ [°] | 105.682(5) |
| $\gamma$ [°] | 90 |
| Volume [Å$^3$] | 642.22(6) |
| Z | 2 |
| $\rho_{calc}$ [gcm$^{-3}$] | 1.672 |
| $\mu$ [mm$^{-1}$] | 0.310 |
| $F(000)$ | 340.0 |
| $2\theta$ range [°] | 4.616 to 61.328 |
| Reflns collected | 7210 |
| Indep reflns ($R_{int}$) | 1751 (0.0188) |
| GOF | 1.095 |
| $R_1$[a], $wR_2$[b] [$I>2\sigma(I)$] | 0.0317, 0.0929 |
| $R_1$, $wR_2$ [all data] | 0.0351, 0.0948 |
| $\Delta\rho$[c] [eÅ$^{-3}$] | 0.26/-0.37 |

[a] $R_1 = \Sigma||F_o| - |F_c||/\Sigma|F_o|$.
[b] $wR_2(F^2) = [\Sigma w(F_o^2 - F_c^2)^2/\Sigma wF_o^4]^{1/2}$.
[c] Maximum and minimum residual electron density.

**Table S2** Comparison of glycine molecular dihedral angle ($\omega_{O-C-C-N}$) between experimental and optimized structures.

|  | N-TGS | P-TGS | TGS-340 K |
|---|---|---|---|
| Experiment | -19.19° | 19.68° | 0° |
| Optimization | -22.16° | 22.17° | 0° |